# FLUX PINNING BY NANO PARTICLES EMBEDDED IN POLYCRYSTALLINE Y-123 SUPERCONDUCTORS

[1]SUSHANT GUPTA, [2]R.S. YADAV, [3]B. DAS

[1, 2, 3]Department of Physics, University of Lucknow, Lucknow–226007, India

Email: sushant1586@gmail.com, yadavramshewak247@yahoo.co.in, bdas226010@gmail.com

## ABSTRACT

Bulk superconductor samples of $YBa_2Cu_{3-x}Zn_xO_{7-\delta}$ with x = 0, 0.01, 0.03 are synthesized by solid-state reaction route. The structural characterisation of all samples has been carried out by x-ray-diffraction (XRD) and transmission electron microscopy (TEM) techniques. The x-ray diffraction patterns indicate that the gross structure/phase of $YBa_2Cu_{3-x}Zn_xO_{7-\delta}$ do not change with the substitution of Zn up to x=0.03. In TEM investigations of Zn-doped Y-based cuprates a number of ZnO nano-flower and nano-rod of Y-211 phase are found dispersed in regular $YBa_2Cu_{3-x}Zn_xO_7$ matrix. These dispersed nano-flowers of ZnO and nano-rods of Y-211 phase may serve as flux-pinning centers. These pinning centers enhance critical current density (Jc) value of these HTSC samples.

## 1. INTRODUCTION

High critical temperature (high-Tc) super-conductors have been the subject of extensive studies carried out to improve their superconducting properties. After the critical transition superconducting temperature (Tc), the critical current density (Jc) is the most important parameter for potential applications of high-Tc superconductors at liquid-nitrogen temperature under an applied magnetic field. The improvement of Jc and its behavior under a magnetic field Jc(H) can be obtained by introducing efficient pinning centers with a size (matching the coherence length) that can suppress the flux flow.

Various methods have been used to introduce effective artificial pinning centers and to increase their number [1-31]. Flux pinning by non-superconducting inclusions and crystal defects is well-documented for $YBa_2Cu_3O_{7-\delta}$ (Y-123 or YBCO for brevity) [1-4]. Many works have indicated that twin boundaries [5, 6], dislocations and stacking faults [1, 7] are possible pinning centers.

Precipitates dispersed into the superconducting matrix are often found to be an effective pinning source. Such precipitates can be obtained by precipitation from a single phase superconductor or by introduction of secondary phase [8-19].

The embedding of nanometre-sized particles [20-26] into the superconducting matrix as flux pinning centres was the goal for a long time since the coherence length of the high-*T*c superconductors (Y-123) is so small (20nm to 300nm).

Nano-sized non-superconducting regions can also be introduced into YBCO bulk superconductors by chemical substitution of atoms in the Y123 lattice (mainly the Cu atoms in the CuO chains [27] or the $CuO_2$ planes [28-30]). These artificially created pinning centres can be effective for enhanced flux pinning in YBCO at intermediate magnetic fields and high temperatures.

The central aim of the present work is to investigate the structural and microstructural changes due to substitution of Zn at Cu site of Y-123 and their possible correlation with superconducting properties (such as critical current density).

## 2. EXPERIMENTAL DETAILS

Samples with nominal composition $YBa_2Cu_{3-x}Zn_xO_{7-\delta}$ (where x = 0, 0.01, 0.03) were synthesised by standard solid state reaction method. The appropriate ratio of the constituent oxides or carbonate i.e. $Y_2O_3$ (99.9%, Alfa Aesar), CuO (99.99%, Alfa Aesar), ZnO (99.0%, Alfa Aesar) and $BaCO_3$ (99.0%, Alfa Aesar) were thoroughly mixed and ground for several hours (2 to 4 hrs) with the help of mortar and pestle. After regrinding and mixing, the powder was kept in a platinum or alumina crucible and heated (calcined) at $875^0C$. The calcinations step served to decompose the carbonate and starting material to interdiffuse for phase formation and chemical homogeneity. After calcinations the material was again ground to subdivide any aggregated products and to further enhance chemical homogeneity. These steps were repeated 3 to 4 times for better homogeneity and phase purity.

The homogeneous powder thus formed was converted into form of pellets before sintering. For this we employed the most widely used technique i.e. dry pressing, which consists of filling a die with powder and pressing at 400 Kg/cm$^2$ into a compacted disc shape. In this way several pellets with varying thickness (1mm to 3mm) were prepared. Finally these pellets were put into alumina crucibles and sintered at about 920±5$^0$C in air. The heating rate to the sintering temperature was about 100$^0$C/hour. After sintering at 920±5$^0$C final annealing was carried out in oxygen atmosphere at partial pressure 10$^{-1}$ atm, at temperature 570$^0$C for 14 hrs in order to maximize the incorporation of oxygen. The annealing temperature is compromise between higher temperatures where oxygen diffusion rates are higher and lower temperatures where the thermodynamic equilibrium favours higher oxygen contents.

Some important factors that can determine 'sample quality' during sample preparation are the stoichiometry, the mixing procedure, the sintering temperature, the oxygen flow during annealing and the rate of sample cooling. If the sample preparation is done with care, a better understanding of the roles these major factors play should ensure that final samples are of high quality.

The structure and phase purity of the powder sample ground from sintered pellets were examined by x-ray diffraction technique using a powder x-ray diffractometer (18 KW, Rigaku Japan) with CuK$_\alpha$ radiation, λ = 1.5408Å. The diffraction data were collected over the diffraction angle range of 2θ= 0-90$^0$ by step scanning with a scanning rate 2$^0$/minute. The structural/ microstructural characteristics was explored by transmission electron microscope (TEM, Tecnai 20 $^2$G) in both the imaging and diffraction modes.

Finally in order to see the effect of Zn doping on the physical properties, the transport critical current density (J$_c$) value has been measured by standard four-probe method as potential difference of 1μv/cm appears across the sample by increasing current at temperatures ranging from 77 to 60 K.

## 3. RESULTS & DISCUSSIONS

Fig. 1 shows the powder x-ray diffraction patterns of YBa$_2$Cu$_{3-x}$Zn$_x$O$_{7-\delta}$ with x = 0, 0.01 and 0.03. The powder diffraction result of these samples showing major phase of Y-123. All the peaks were indexed on the basis of Y-123 structure. The crystal structure was found to be orthorhombic (Pmmm space group) with Zn substitution up to x=0.03.

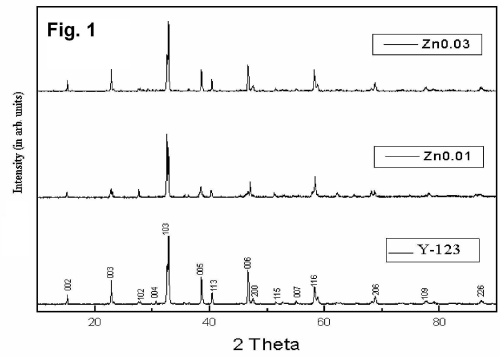

**Fig. 1: X-ray powder diffraction patterns of YBa$_2$Cu$_{3-x}$Zn$_x$O$_{7-\delta}$ system with x = 0.00, 0.01, 0.03.**

The Zero-resistance critical temperature Tc (R=0) decreases systematically with the increase in concentration of zinc. The values of Tc (R=0) for the samples with zinc concentrations x= 0, 0.01, and 0.03 are 91K, 60K and 55K respectively.

In order to explore the effect of Zn doping on the microstructural characteristics in YBa$_2$Cu$_{3-x}$Zn$_x$O$_{7-\delta}$ superconductors, transmission electron microscopy was employed in imaging and diffraction modes.

In TEM investigation of Zn-doped Y-based cuprates a number of ZnO nano-flower are found dispersed in regular YBa$_2$Cu$_{3-x}$Zn$_x$O$_7$ matrix as shown in Figure 2. These dispersed nano structure may act as flux pinning centers that result high transport (inter-grains) critical current density J$_c$ ~ 10$^3$A / cm$^2$ at 50K.

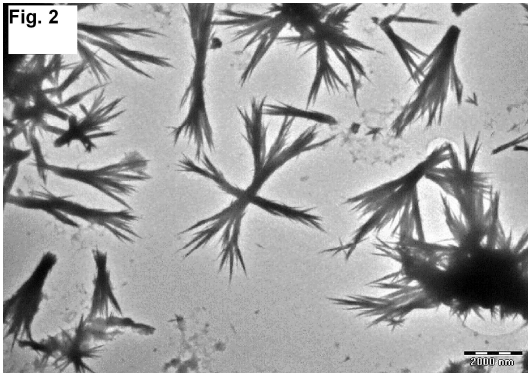

**Fig. 2: Transmission electron micrograph of Zn-doped Y-123 superconductors showing several nano-flowers.**

As the most prominent pinning center in melt textured high-$T_c$ superconductors is the insulating $Y_2BaCuO_5$ (Y-211) particle being created during the growth of YBCO and providing flux pinning. Figure 3(a) shows, a nano-rod of this Y-211 phase and its selected area diffraction (SAD) pattern is shown in Figure 3(b).

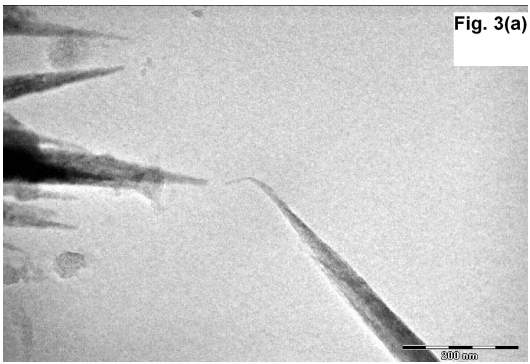

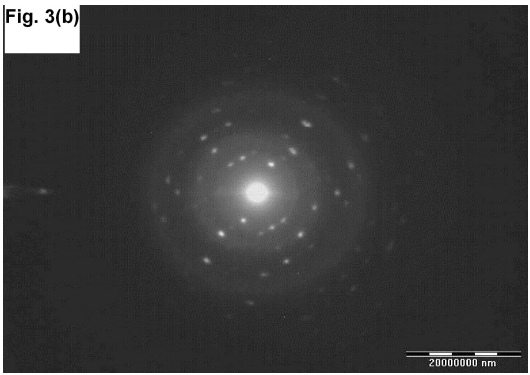

**Fig. 3(a) TEM of Zn doped Y-123 showing nano-rod of Y-211 phase 3(b) its corresponding SAD pattern along [110] direction.**

The variation of the transport critical current density (Jc) of $YBa_2Cu_{3-x}Zn_xO_{7-\delta}$ with different Zinc concentration revealed one order enhancement in Jc value as shown in fig. 4. In Zinc free sample critical current density (Jc) has been measured to be about $1.5 \times 10^2$ A/cm$^2$ but Jc=$2.57 \times 10^3$ A/cm$^2$ has been found for zinc doped (x=0.03) sample.

A serious handicap of the superconducting cuprates is the easy motion of their vortices, so that their critical current density tends to decrease dramatically when the temperature approaches Tc or when a high magnetic field is applied. People speculated that one could improve the critical current in crystals by introducing defects. A magnetic field penetrates a type-II superconductor in the form of vortices. Each vortex, carrying a flux quantum Φ=h/2e, consists of a cylindrical core of radius ξ, the coherence length of the material, and a current circulating around the core out to a distance λ, the material penetration depth. When a current flows in the superconductor, the Lorentz force produces energy dissipation and consequently the disappearance of superconductivity.

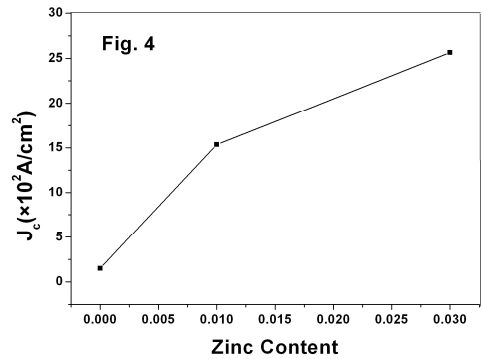

**Fig. 4: The variation of transport critical current density Jc versus Zinc content.**

The mobility of the vortices can be minimized by introducing efficient pinning centers that pin the vortices. Thus the critical current density Jc depends on flux pinning of vortices. Nano-defects will be the best pinning centers since high Tc superconductor of type Y-123, Tl-2212 and Tl-2223 have a coherent length of range 20nm to 300nm. In present Zn doped Y-based system some nano-flowers of ZnO and nano-rods of Y-211 phase have already been observed in electron microscopic investigations. These nano-flowers and nano-rods may act as effective flux pinning centers. These pinning centers enhance critical current density (Jc) value of these HTSC materials as observed in our transport Jc measurements.

## 4. CONCLUSION

We have successfully prepared samples of type YBa$_2$Cu$_{3-x}$Zn$_x$O$_{7-\delta}$ with x = 0, 0.01 and 0.03 by standard ceramic method. The partial replacement of Cu by Zn does not affect the orthorhombic structure of the Y-123 phase. The structural/microstructural characteristics were explored through transmission electron microscopy (TEM). An interesting result has been found in TEM study that the dispersion of nano-flowers and nano-rod of ZnO and Y-211 phase respectively in local region of Zn-doped Y-based cuprates. These dispersed nano-flowers of ZnO and nano-rods of Y-211 phase possibly behave as flux pinning centers. The critical current density Jc increases as Zinc content increases as a result of the flux pinning centers of nano-flowers of ZnO and nano-rods of Y-211 phase.

## REFERENCES

[1] M. Murakami, "Processing of bulk YBaCuO", *Supercond. Sci. Technol.*, Vol. 5, 1992, pp. 185-203.

[2] D. Shi, K. C. Goretta, J. G. Chen and A. C. Biondo, "Critical current and flux pinning by crystal defects in melt textured YBa$_2$Cu$_3$O$_x$", *Ceram. Trans.*, Vol. 18, 1991, pp. 373-382.

[3] V. Selvamanickam, M. Mironova, S. Son and K. Salama, "Flux pinning by dislocations in deformed melt textured YBa$_2$Cu$_3$O$_x$ superconductors", *Physica C*, Vol. 208, 1993, pp. 238-244.

[4] V. Pavate, L.B. Williams, E.P. Kvam, G. Kozlowski, W. Endres and C.E. Oberly, "Identification and correlation of microstructural defects with flux pinning in Ni-doped melt textured YBa$_2$Cu$_3$O$_{7-\delta}$", *Appl. Phys. Lett.*, Vol. 65, 1994, pp. 246-248.

[5] L.J. Swartzendruber, D.L. Kaiser, F.W. Gayle, L.H. Bennett and A. Roytburd, "Low field flux pinning in twinned and detwinned single crystals of YBa$_2$Cu$_3$O$_{7-x}$", *Appl. Phys. Lett.*, Vol. 58, 1991, pp. 1566-1568.

[6] H. Fujimoto, T. Tahuchi, M. Murakami and N. Koshizuka, "The effect of twin boundaries on the flux pinning in MPMG processed YBCO", *Physica C*, Vol. 211, 1993, pp. 393-403.

[7] J. Ringnalda, C. Kiely, P. Fox and G.J. Tatlock, "Stacking fault structures in melt-processed YBa$_2$Cu$_3$O$_{7-\delta}$ superconductors", *Phil. Mag. A,* Vol. 69, 1994, pp. 729-739.

[8] S. Sengupta, D. Shi, Z. Wang, A.C. Biondo, U. Balachandran and K.C. Goretta, "Effect of Y$_2$BaCuO$_x$ precipitates on flux pinning in melt-processed YBa$_2$Cu$_3$O$_x$", *Physica C*, Vol. 199, 1992, pp. 43-49.

[9] D. Shi, S. Sengupta, J.S. Luo, C. Varanasi and P.J. McGinn, "Extremely fine precipitates and flux pinning in melt-processed YBa$_2$Cu$_3$O$_x$", *Physica C*, Vol. 213, 1993, pp. 179-184.

[10] S. Sengupta, D. Shi, J. S. Luo, A. Buzdin, V. Gorin, V. R. Todt, C. Varanasi and P. J. McGinn, "Effect of extremely fine Y$_2$BaCuO$_5$ precipitates on the critical current density of melt-processed YBa$_2$Cu$_3$O$_x$", *J. Appl. Phys.*, Vol. 81, 1997, pp. 7396-7408.

[11] N. Hari Babu, E.S. Reddy, D.A. Cardwell, A.M. Campbell, C.D. Tarrant and K.R. Schneider, "Artificial flux pinning centers in large, single-grain (RE)-Ba-Cu-O superconductors" *Appl. Phys.Lett.*, Vol. 83, 2003, pp. 4806-4808.

[12] N. Hari Babu, E. Sudhakar Reddy, D.A. Cardwell and A.M. Campbell, "New chemically stable, nano-size artificial flux pinning centres in (RE)–Ba–Cu–O superconductors" *Supercond. Sci. Technol.*, Vol. 16, 2003, p. L44.

[13] N. Hari Babu, K. Iida and D.A. Cardwell, "Enhanced magnetic flux pinning in nano-composite Y-Ba-Cu-O superconductors", *Physica C*, Vol. 445, 2006, pp. 353-356.

[14] M. Murakami, S. Gotoh, H. Fujimoto, K. Yamaguchi, N. Koshizuka and S. Tanaka, "Flux pinning and critical currents in melt processed YBaCuO superconductors", *Supercond. Sci. Technol.*, Vol. 4, 1991, pp. S43-50.

[15] N. Hari Babu, M. Kambara, Y. Shi, D.A. Cardwell, C.D. Tarrant and K.R. Schneider, "The chemical composition of uranium-containing phase particles in U-doped Y–Ba–Cu–O melt processed superconductor" *Physica C*, Vol. 392, 2003, pp. 110-115.

[16] S. Jin, G. W. Kammlott, T. H. Tiefel, T.T. Kodas, T. L. Ward and D. M. Kroeger, "Microstructure and properties of the Y-Ba-Cu-O superconductor with submicron "211" dispersions" *Physica C.,* Vol. 181, 1992, pp. 57-62.

[17] M. Chopra, S.W. Chan, R. L. Meng and C.W. Chu, "$Y_2BaCuO_5$ addition and its effects on critical currents in large grains of $YBa_2Cu_3O_{7-\delta}$: A quantitative microstructural study", *J. Mater. Res.*, Vol. 11, 1996, pp. 1616-1626.

[18] C. Meingast, P.J. Lee and D.C. Larbalestier, "Quantitative description of a high Jc Nb-Ti superconductor during its final optimization strain, microstructure, Tc, Hc2, and resistivity", *J. Appl. Phys.*, Volume: 66, 1989, pp. 5962–5970.

[19] K. Zmorayova, P. Diko, A.E. Carrillo, F. Sandiumenge and X. Obradors, "Quantitative analysis of Y2BaCuO5 particle distribution in a melt-grown YBa2Cu3O7/Y2BaCuO5 bulk superconductor", *Supercond. Sci. Technol.*, Vol. 18, 2005, pp.948-952.

[20] T. Haugan, P. N. Barnes, R. Wheeler, F. Meisenkothen and M. Sumption, "Addition of nanoparticles dispersions to enhance flux pinning of the $YBa_2Cu_3O_{7-x}$ superconductor", *Nature,* Vol. 430, 2004, pp. 867-870.

[21] S. Kang, A. Goyal, J. Li, A. A. Gapud, P. M. Martin, L. Heatherly, J. R. Thompson, D. K. Christen, F. A. List, M. Paranthaman and D. F. Lee, "High-Performance High-$T_c$ Superconducting Wires", *Science*, Vol. 311, 2006, pp. 1911-1914.

[22] A. Goyal, S. Kang, K.J. Leonard, P.M. Martin, A.A. Gapud, M. Varela, M. Paranthaman, A.O. Ijaduola, E.D. Specht, J. R. Thompson, D.K. Christen, S.J. Pennycook and F.A. List, "Irradiation-free, columnar defects comprised of self-assembled nanodots and nanorods resulting in strongly enhanced flux-pinning in $YBa_2Cu_3O_{7-\delta}$ films", *Supercond. Sci. Technol.*, Vol. 18, 2005, pp. 1533–1538.

[23] N. Hari Babu*,* K Iida and D.A. Cardwell, "Flux pinning in melt-processed nanocomposite single-grain superconductors", *Supercond. Sci. Technol.*, Vol. 20, 2007, pp. S141–S146.

[24] S. Kang, K.J. Leonard, P.M. Martin, J. Li and A. Goyal, "Strong enhancement of flux pinning in $YBa_2Cu_3O_{7-\delta}$ multilayers with columnar defects comprised of self-assembled BaZrO3 nanodots", *Supercond. Sci. Technol.*, Vol. 20, 2007, pp. 11-15.

[25] T. J. Haugan *et al.*, "Island growth of $Y_2BaCuO_5$ nanoparticles in (211~1.5nm/ 123~10nm) xN composite multilayer structures to enhance flux pinning of $YBa_2Cu_3O_{7-\delta}$ films", *J. Mater. Res.*, Vol. 18, 2003, pp. 2618–2623.

[26] P.N. Barnes, T.J. Haugan, F.J. Baca, C.V. Varanasi, R. Wheeler, F. Meisenkothen and S. Sathiraju, "Inducing self-assembly of $Y_2BaCuO_5$ nanoparticles via Ca-doping for improved pinning in $YBa_2Cu_3O_{7-x}$", *Physica C*, Vol. 469, 2009, pp. 2029-2032.

[27] Y. Ishii, J. Shimoyama, Y. Tazaki, T. Nakashima, S. Horii and K. Kishio, "Enhanced flux pinning properties of $YBa_2Cu_3O_y$ by dilute impurity doping for CuO chain", *Appl. Phys. Lett.*, Vol. 89, 2006, p. 202514.

[28] G. Krabbes, G. Fuchs, P. Schatzle, S. Gruss, J.W. Park, F. Hardinghaus, G. Stover, R. Hayn, S-L. Drechsler and T. Fahr, "Zn doping of $YBa_2Cu_3O_7$ in melt textured materials: peak effect and high trapped fields" *Physica C,* Vol. 330, 2000, pp. 181-190.

[29] L. Shlyk, G. Krabbes, G. Fuchs, G. Stover, S. Gruss and K. Nenkov, "Pinning behavior and magnetic relaxation in melt-processed YBCO doped with Li, Ni and Pd" *Physica C*, Vol. 377, 2002, pp. 437-444.

[30] V. Antal, M. Kanuchova, M. Sefcikova, J. Kovac, P. Diko, M. Eisterer, N. Horhager, M. Zehetmayer, H.W. Weber and X. Chaud, "Flux pinning in Al doped TSMG YBCO bulk superconductors", *Supercond. Sci. Technol.*, Vol. 22, 2009, p. 105001.

[31] L. Civale, "Vortex pinning and creep in high-temperature superconductors with columnar defects", *Supercond. Sci. Technol.*, Vol. 10, 1997, pp. A11-A28.